\documentclass{emulateapj}

\slugcomment{Accepted for publication in  ApJ {\it Letters}}

\newcommand\chandra{{\it Chandra}}

\newcommand\rxte{{\it RXTE}}
\newcommand\xmm{{\it XMM-Newton}}

\newcommand\s{{\rm~s}}
\newcommand\ks{{\rm~ks}}
\newcommand\mhz{{\rm~mHz}}
\newcommand\mpc{{\rm~Mpc}}

\newcommand\hz{{\rm~Hz}}
\newcommand\kev{{\rm~keV}}
\newcommand\ev{{\rm~eV}}
\newcommand\kms{\ifmmode {\rm~km\ s}^{-1} \else ~km s$^{-1}$\fi}
\newcommand\Hunit{\ifmmode {\rm~km\ s}^{-1}\ {\rm Mpc}^{-1}
        \else ~km s$^{-1}$ Mpc$^{-1}$\fi}
\newcommand\ctssec{\ifmmode {\rm~count\ s}^{-1} \else ~count s$^{-1}$\fi}
\newcommand\ergsec{\ifmmode {\rm~erg\ s}^{-1} \else
        ~erg s$^{-1}$\fi}
\newcommand\funit{\ifmmode {\rm~erg\ s}^{-1}\;{\rm cm}^{-2} \else
        ~ergs s$^{-1}$ cm$^{-2}$\fi}
\newcommand\phflux{\ifmmode {\rm~photon\ s}^{-1}\;{\rm cm}^{-2}
        \else   ~photon s$^{-1}$ cm$^{-2}$\fi}
\newcommand\efluxA{\ifmmode {\rm~erg\ s}^{-1}\;{\rm cm}^{-2}\;{\rm
        \AA}^{-1} \else ~erg s$^{-1}$ cm$^{-2}$ \AA$^{-1}$\fi}
\newcommand\efluxHz{\ifmmode {\rm~erg\ s}^{-1}\;{\rm cm}^{-2}\;{\rm
        Hz}^{-1} \else ~erg s$^{-1}$ cm$^{-2}$ Hz$^{-1}$\fi}
\newcommand\cc{\ifmmode {\rm~cm}^{-3} \else cm$^{-3}$\fi}
\newcommand\FWHM{\ifmmode {\rm~FWHM} \else ${\rm~FWHM}$\fi}
\newcommand\Msun{\ifmmode M_{\odot} \else $M_{\odot}$\fi}
\newcommand\Lsun{\ifmmode L_{\odot} \else $L_{\odot}$\fi}

\newcommand\hbeta{\ifmmode {\rm H}\beta \else H$\beta$\fi}
\newcommand\Kalpha{\ifmmode {\rm K}\alpha \else K$\alpha$\fi}
\newcommand\nh{\ifmmode N_{\rm H} \else N$_{\rm H}$\fi}
\usepackage{graphicx}
\usepackage{here}

\begin{document}

\title{Black Hole Mass of the Ultraluminous X-ray source M82 X--1}

\author{Gulab C. Dewangan\altaffilmark{1}, Lev
  Titarchuk\altaffilmark{2} \& Richard E. Griffiths\altaffilmark{1}}
\altaffiltext{1}{Department of Physics, Carnegie Mellon University,
  5000 Forbes Avenue, Pittsburgh, PA 15213 USA; {\tt email: gulabd@cmu.edu} }
\altaffiltext{2}{Center for Earth Observing and Space Research, George
  Mason University, Fairfax, VA 22030; and US Naval Research
  Laboratory, Code 7655, Washington, DC 20375-5352}

\begin{abstract}
  We report the first clear evidence for the simultaneous presence of a
  low frequency break and a QPO in the fluctuation power spectrum of a well
  known ultraluminous X-ray source (ULX) in M82 using long \xmm{}
  observations. The break occurs at a frequency of $34.2_{-3}^{+6}\mhz$.
  The QPO has a centroid at $\nu_{QPO} = 114.3\pm1.5\mhz$, a coherence $Q
  \equiv \nu_{QPO}/\Delta\nu_{FWHM} \simeq 3.5$ and an amplitude (rms) of
  $19\%$ in the $2-10\kev$ band. The power spectrum is approximately
  flat below the break frequency and then falls off above the break 
frequency as a power law
  with the QPO superimposed. This form of the power spectrum is
  characteristic of the Galactic X-ray binaries (XRBs) in their high
  or intermediate states. M82 X-1 was likely in an intermediate state
  during the observation. The EPIC PN spectrum is well described by a
  model comprising  an absorbed power-law ($\Gamma \sim 2$) and an
  iron line at $\sim 6.6\kev$ with a width $\sigma \sim 0.2\kev$ and
  an equivalent width of $\sim 180\ev$.  Using the well established
  correlations between the power and energy spectral parameters for
  XRBs, we estimate a black hole mass 
for M~82 X-1 in the range of $\sim 25 -
  520M\odot$ including systematic errors that arise due to the
 uncertainty in the calibration of the photon spectral index versus QPO
  frequency relation.  
\end{abstract}


\section{Introduction}
Ultra-luminous X-ray sources (ULXs) are extra-nuclear point X-ray
sources with luminosities exceeding $\sim 10^{39}{\rm~erg~s^{-1}}$.
These objects were first discovered with {\it Einstein} observations
(reviewed by Fabbiano 1989). 
The nature of ULXs
continues to be an enigma, since their adopted isotropic high energy
output surpasses the Eddington limit of even the most massive stellar
mass black holes (BHs), sometimes by large factors.
Several models have been proposed to explain the high luminosities of
ULXs. The most popular is the  ``intermediate mass
black hole (IMBH)'' with mass $M_{BH} \simeq 10^2 -
10^4{\rm~M_{\odot}}$ (e.g., Colbert \& Mushotzky 1999, hereafter CM99) bridging
the gap between stellar mass BHs in XRBs and
super-massive BHs in active galactic nuclei \footnote{In CM99
  the BH mass was evaluated using the best-fit parameters
  of the {\tt XSPEC} BMC model (see Shrader \& Titarchuk 1999
 for details)}.  
Other popular models include XRBs with
anisotropic emission (King et al. 2001), beamed XRBs with relativistic
jets directly pointing towards us i. e., scaled down versions of
blazars (Mirabel \& Rodriguez 1999), and XRBs with super-Eddington
accretion rates (Begelman 2002).  Several observations suggest that
ULXs may be similar or scaled up versions of XRBs. Discovery of orbital modulations from several ULXs
(Bauer et al. 2001; Sugiho et al. 2001) implies their binary nature.  Recent
\chandra{} and \xmm{} observations of ULXs show soft X-ray excess
emission which has been interpreted as the emission from accretion
disks with temperatures in the range  $\sim 100 -500\ev$ (see
Shrader \& Titarchuk 2003 and review by Miller \& Colbert 2003 and
references therein).  Observations of spectral transitions between
low/hard and high/soft states in two ULXs in IC~342 (Kubota et al.
2001) further demonstrate their similarity with the XRBs. 
The recent observation of a break at a frequency of $\sim 2.5\mhz$  in 
the power density spectrum (PDS) of the ULX in NGC~5408 
suggests a BH mass of $\sim 100M\odot$ (Soria et al. 2004).  

The enigmatic nature of ULXs can be understood by determining their BH
masses.  The most reliable method is to measure the mass function
through the secondary mass and  orbital parameters such as
velocities, period, orbit size etc., which can be measured only if the
secondary is optically identified. There are two cases of optical 
identification (Kuntz et al. 2005; Liu et al. 2004); however, 
the orbital parameters of 
any ULX binary system are yet to be measured. Good progress can still be made
by establishing a direct physical connection between ULXs and XRBs.
This can be done by comparing the characteristic time scales such as
those associated with the low frequency quasi-periodic oscillations
(QPOs). However, this comparison is ambiguous without the information
about a low frequency break, the shape of the PDS and
energy spectrum.

M82 X-1 is one of the brightest ULX (CXO M82 J095550.2+694047, source 7 
in Matsumoto et al.
2001) that is well suited for a determination of the shape of the PDS.
\chandra{} High Resolution Camera observations showed that the source
is unresolved and off-nuclear (Matsumoto et al.  2001). The ULX has no
optically bright counterpart (Kaaret et al.  2001): there is a likely 
IR counterpart which is a super star cluster (Kaaret et al. 2004). Recently,
Strohmayer \& Mushotzky (2003), hereafter SM03, discovered a QPO in
M82 X-1 at a frequency of $54{\rm~mHz}$ based on a $27{\rm~ks}$ \xmm{}
observation. They also found QPOs in the $50-100{\rm~mHz}$ frequency
range in the \rxte{} data.
Fiorito \& Titarchuk (2004), hereafter FT04 apply a new method to
determine the BH mass of M82 X-1.  The method uses the spectral index-QPO
low-frequency correlation that has been recently established in
BH XRBs GRS 1915+105, XTE J1550-564, 4U 1630-47, and
others (Titarchuk \& Fiorito 2004, hereafter TF04) .  Using scaling
arguments and the correlation derived from 
Galactic BHs, they conclude that M82 X-1 is an intermediate mass BH
 of order 1000 M$_{\odot}$.
We revisit the BH mass estimate in M82~X-1 using new data 
on the power spectrum and energy spectra.
    
\section{Observation, Data Reduction \& Analysis}
M82 was observed using \xmm{} on 2001 May 5 for $30\ks$ and again on 2004 April
21 for $103\ks$. The first observation led to the discovery of a
$54\mhz$ QPO from the ULX (SM03). Here we consider the EPIC data from
the second observation only. The EPIC PN and MOS cameras were operated
in the full frame mode using the medium filter. We used  the SAS
version 6.1 and the most recent calibration data base to
process and filter the event data. Examination of the background rate
above $10\kev$ showed that the observation is completely swamped by
the particle background after an elapsed time of $71.5\ks$ and  this
latter period was therefore
excluded from the rest of the analysis.

In the central regions of M82, \chandra{} images revealed diffuse emission
and several point sources, M82 X-1 being the brightest among them
(Matsumoto et al. 2001). The diffuse emission is evident in the EPIC
images but the point sources are not resolved. Following SM03, we
extracted PN and MOS events using a $18\arcsec$ circular region around
the bright point source whose position is consistent with the
\chandra{} position of M82 X-1. Above $2\kev$, the diffuse emission
contributes only $< 10\%$ to the point source inside the $18\arcsec$
circular region. At the faintest flux level of M82 X-1, nearby point
sources contribute $\la 30\%$ to the $0.5-10\kev$ X-ray emission in
the $18\arcsec$ region (Matsumoto et al. 2001; SM03).
\subsection{The power spectrum}
For temporal analysis, we combined the PN and MOS data and used the
continuous exposure of $70.2\ks$ during which both the PN and MOS
cameras operated simultaneously. We calculated a power density
spectrum (PDS) using the background corrected PN+MOS light curves
sampled at $0.5\s$.  Figure~\ref{f1} ({\it left}) shows the $2-10\kev$
PDS of M82 X-1 rebinned by a factor of $1024$ yielding a frequency
resolution of $7.8\mhz$. The PDS continuum is approximately flat at
low frequencies below $\sim 30\mhz$ and then falls off approximately
following a power law up to $\sim 200\mhz$ where the white noise
arising from the poisson errors starts to dominate the PDS.  There is
a prominent QPO with its peak frequency near $114\mhz$.

We fitted the PDS with two models: ($i$) a broad Lorentzian for the
continuum and a narrow Lorentzian for the QPO, and ($ii$) a broken
power law (BPL) for the continuum and a Lorentzian (L) for the QPO.
We also used a constant to account for the poisson noise.  Both models
provided statistically acceptable fits, giving a minimum $\chi^2$ of
$128.6$ for 121 degrees of freedom (dof) and $127.9$ for 120 dof for the
double Lorentzian and BPL$+$L models, respectively. The errors on the
best-fit PDS model parameters, quoted below, are at a $1\sigma$ level.
The double Lorentzian model resulted in a QPO centroid frequency
$\nu_{QPO} = 114.6\pm1.5\mhz$, a width $\nu_{FWHM} =
31.3_{-2.5}^{+2.9}\mhz$, an amplitude $A_{QPO}=0.037.4\pm0.0024$ and
the broad Lorentzian centroid $\nu_{0} = 11.4_{-4.4}^{+4.2}\mhz$, a
width $\nu_{FWHM} = 60.2_{-6.3}^{+7.5}\mhz$ and an amplitude $A =
0.038\pm0.0034$. The BPL$+$L model resulted in a QPO centroid frequency
$\nu_{QPO} = 114.3\pm1.5\mhz$, a width $\nu_{FWHM} =
32.7_{-2.6}^{+2.9}\mhz$, an amplitude $A_{QPO}=0.038\pm0.0024$, and
 broken power-law indices of $\Gamma_1 = 0.11_{-0.09}^{+0.10}$
below the break frequency $\nu_b = 34.2_{-2.9}^{+5.7}\mhz$ and
$\Gamma_2 = 2.32_{-0.39}^{+0.58}$ above the break frequency, and an
amplitude $A_{BPL} = 0.36\pm0.03$ at a reference frequency of
$10\mhz$. The total integrated power ($0.001-1\hz$) and the QPO power
expressed as $rms/mean$ are $23\%$ and $19\%$, respectively.  The use
of a broken power law is an improvement ($\Delta \chi^2 = -24.8$ for
two additional parameters) over a simple power law at a statistical
significance level of $>99.99\%$ based on the maximum likelihood ratio
test.  Thus the break in the continuum of the power spectrum is real.
The best-fit BPL$+$L model is shown in Fig.~\ref{f1}({\it left}) as a
thick line.

 For comparison we have also calculated the PDS of an XRB
XTE~J1550-564 using the \rxte{} observation of 10 September 1998.
XTE~J1550-564 shows a large range in its low frequency QPO centroid
($\nu_{QPO} \sim 0.08 - 18\hz$; Sobczak et al. 2000a). Our choice of
the particular \rxte{} observation relies on the fact that
XTE~J1550-564 showed a power-law (PL) photon index of $\Gamma \sim 2.0$ on
10 September 1999 (Sobczak et al. 2000b), which is very similar to the
$3-10\kev$ photon index of M82 X-1 (see below). Fig.~\ref{f1}({\it
  right}) shows the PDS of XTE~J1550-564 calculated from the PCA light
curve sampled at $0.125\s$ which was obtained from the \rxte{} public
data archive. We used  a model comprising a broken PL and
two Lorentzians to fit the PDS. This model resulted in a  statistically
acceptable fit (minimum $\chi^2 = 94.2$ for 118 dof). The broken PL
has best-fit parameters: $\Gamma_1 = 0.03\pm0.006$ below the
break frequency $\nu_b = 0.27\pm0.007\hz$, $\Gamma_2 = 0.89\pm0.01$
above the break and an amplitude $A_{BPL} = 0.026\pm0.0005$ at a
reference frequency of $10\mhz$. The two QPOs have centroid
frequencies $\nu_{QPO} = 1.034\pm0.004$ and $2.055\pm0.016\hz$, widths
$\nu_{FWHM} = 0.125\pm0.010$ and $0.39\pm0.05\hz$ and amplitudes
$A_{QPO} = 0.026\pm0.0005$ and $0.0046\pm0.0003$ for the fundamental and first
harmonic, respectively. The total integrated power ($0.01-4\hz$) and the QPO (fundamental) power expressed as $rms/mean$ are $23\%$ and $16\%$, respectively.  
  
\subsection{The energy spectrum}
\begin{figure}
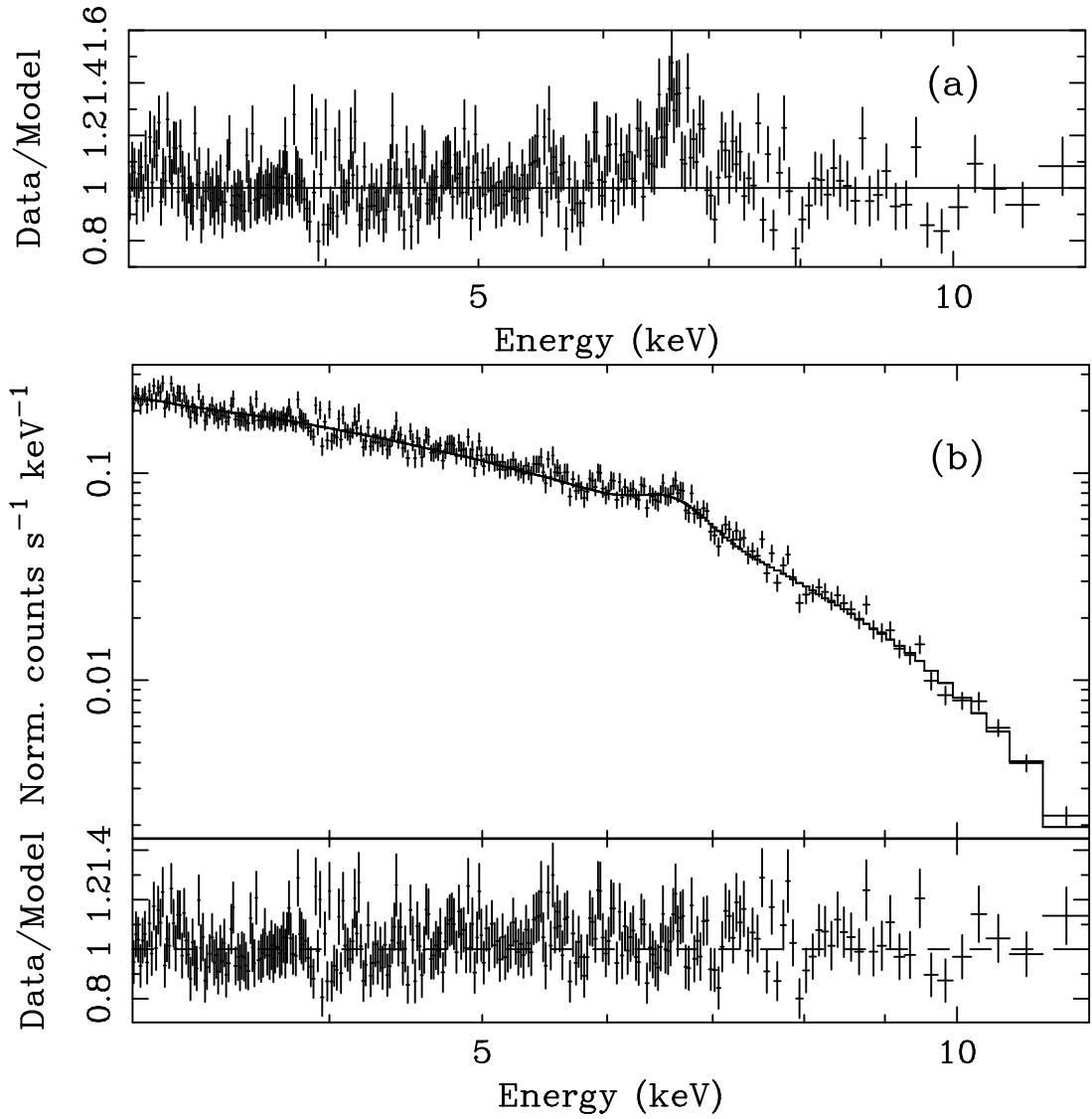

   \centering \includegraphics[angle=-90]{f2a.ps}
   \includegraphics[angle=-90]{f2b.ps}
   \caption{(a) Ratio of EPIC PN data and the best-fit absorbed
     PL model showing the iron K$\alpha$ line. (b) The PN
     spectrum and the best-fit absorbed PL$+$Gaussian line
     model ({\it upper panel}) and their ratio ({\it lower panel}).}
   \label{f2}
 \end{figure}

We extracted PN and MOS spectra using a $18\arcsec$ circular region
centered at the position of M82 X-1. We also extracted PN and MOS
background spectra using nearby circular regions free of sources. We
created appropriate response files using the SAS tasks {\tt rmfgen}
and {\tt arfgen}.  The spectra were grouped to a minimum of $20$
counts per spectral channel and analyzed with {\tt XSPEC 11.3}. The
errors on the best-fit spectral parameters are quoted at a $90\%$
confidence level. An absorbed power law (PL) model fitted to the PN
data showed strong soft excess emission below $\sim 3\kev$ due to the
presence of the diffuse soft X-ray emission in the source extraction
region. We ignored the data below $3\kev$ and performed the spectral
fitting of the PN and MOS data separately. A simple PL model
poorly describes both the PN and MOS data. We show the ratio of the PN
data and the best-fitting absorbed PL model in
Figure~\ref{f2}(a). A prominent iron K$\alpha$ line at $\sim 6.6\kev$
is evident. Addition of a Gaussian line (GL) to the PL model
improved the fit significantly ($\Delta \chi^2 = -71.7$ for three
additional parameters) and resulted in a good fit (minimum $\chi^2 =
898.9$ for 944 dof. The Gaussian line has a centroid energy $E_{line} =
6.61_{-0.08}^{+0.07}\kev$, a width $\sigma = 278_{-115}^{+161}\ev$, a
line flux $f_{line} = 1.9_{-0.6}^{+0.7} \times
10^{-5}{\rm~photons~cm^{-2}~s^{-1}}$ and an equivalent width of $\sim
180\ev$. The best-fit photon index is $2.00_{-0.07}^{+0.07}$ and the
the absorption column is $\nh = 2.5_{-0.6}^{+0.6}\times
10^{22}{\rm~cm^{-2}}$. The observed $2-10\kev$ flux is
$9.6\times10^{-12}{\rm~erg~cm^{-2}~s^{-1}}$ and the corresponding
luminosity is $1.7\times 10^{40}{\rm~erg~s^{-1}}$ assuming a distance
of $3.9\mpc$ (Sakai \& Madore 1999). In Figure~\ref{f2}b, we show the
best-fit PL$+$GL model to the EPIC PN data. The PL$+$GL model fitted
to the MOS data resulted in slightly flatter PL ($\Gamma_X =
1.80_{-0.09}^{+0.11}$), an absorption column
$\nh=1.8_{-0.3}^{+0.7}\times 10^{22}{\rm~cm^{-2}}$ and a weaker iron
line, $EW \sim 94\ev$.

\section{Discussion}
Based on long \xmm{} observations, we show clear evidence for the
simultaneous presence of a low frequency break at $\nu_b
=33.5_{-3.0}^{+6.7}\mhz$ and a QPO with centroid frequency $\nu_{QPO}
= 113.2\pm1.5\mhz$ in the power spectrum of M82 X-1.  The QPO
frequency is similar to that found from the \rxte{} observation of
1997 July 21, while it is about a factor of two larger than that found
in the first \xmm{} observation of 2001 May and an \rxte{} observation
of 1997 February 24 (SM03). Thus the QPO frequency of M82 X-1 is
variable.

\begin{figure}
  \centering \includegraphics[width=8cm]{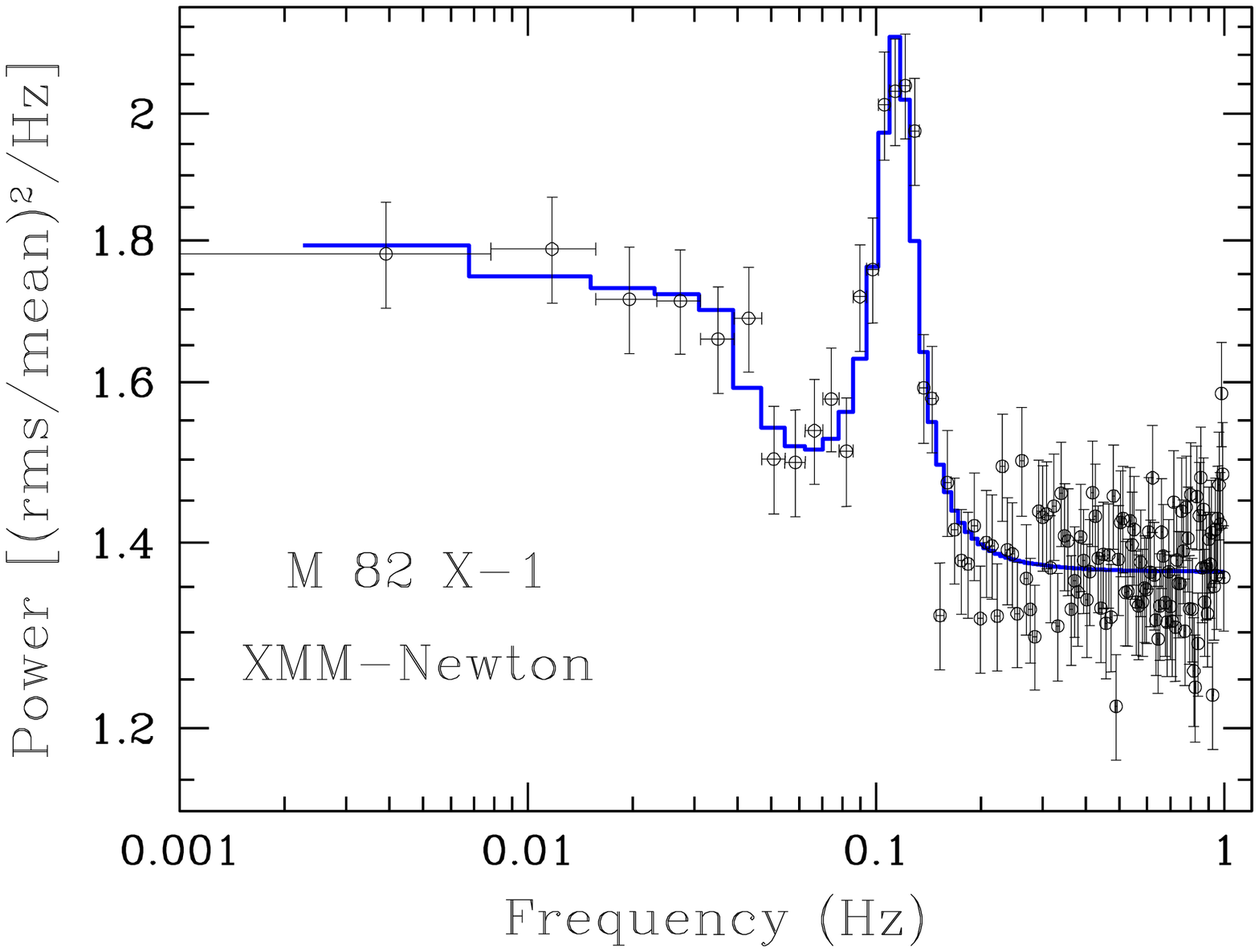}
  \includegraphics[width=8cm]{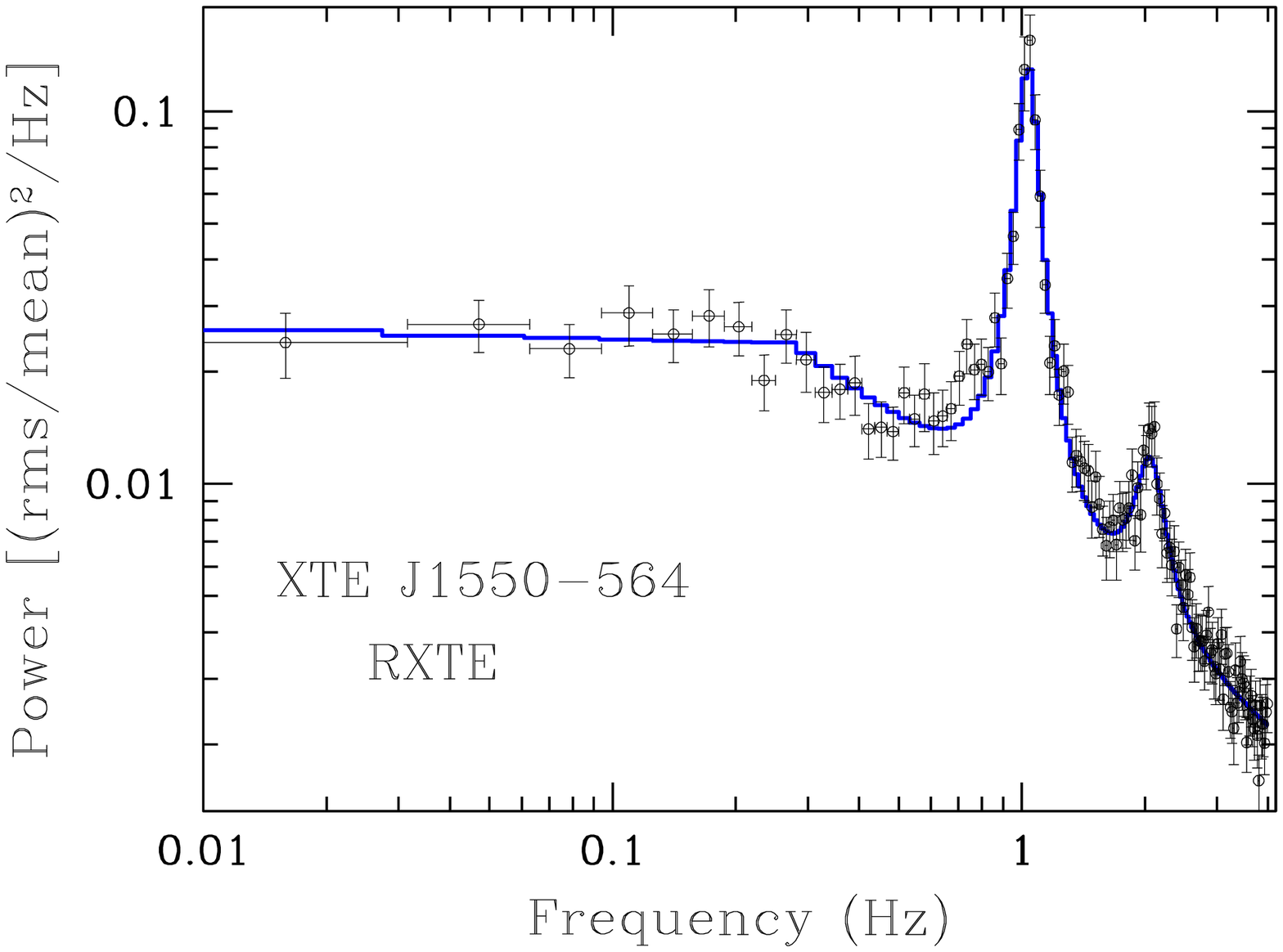}
  \caption{PDS of the ULX M82 X-1 and an XRB
    XTE~J1550-564 having very similar PL photon indices
    ($\Gamma \sim 2.0$). {\it Left:} PDS of M82 X-1 derived
    from the EPIC PN and MOS data above $2\kev$.  The white noise
    level expected  ($\sim 1.4$) from the poisson errors has not been subtracted.
    The frequency resolution is $7.8{\rm~mHz}$. The best-fitting model
    comprising a broken power-law and a Lorentzian is shown as a
    thick line. {\it Right:} PDS of the  XRB
    XTE~J1550-564 derived from the \rxte{} observations of 1998
    September 10.  The best-fit model consisting of a broken power law
    and two Lorentzians is shown as a thick line. }
  \label{f1}
\end{figure}

The M82 X-1 PDS is flat below the break frequency and then  falls off above the break 
as a power law with a prominent QPO on top of it. This form of the PDS
is characteristic of  XRBs in their high/soft state
or intermediate state (e.g., McClintock \& Remillard 2003;  
see also Fig.~\ref{f1} 
for a comparison).  PDSs of
many BH XRBs display low frequency QPOs in the frequency range
of $\sim 0.05 - 30\hz$ (Remillard et al. 2002). The QPO frequency is
strongly correlated with the break frequency (Wijnands \& van der Klis
1999). The break frequency is always lower than the QPO frequency by
at least a factor of three (see, Wijnands \& van der Klis 1999). M82
X-1 follows this correlation as its QPO frequency is about a
factor three larger than the break frequency. {\it Thus the PDS of the ULX is
 very similar to the XRB PDSs in their high/soft or transition
state.}

Since the characteristic time scales of accretion powered sources scale
with BH mass, the similarity of the PDSs of M82 X-1 and XRBs provides
the opportunity to determine the ratio of their BH masses.  However, as
stated above, the XRBs show a large range in their low
frequency QPO and break frequencies. Therefore, the frequency scaling
factor between any two power spectra of two XRBs will lead to
an incorrect determination of the ratio of their BH masses. Fortunately,
the PDS features (break and QPO frequencies) are well correlated with
the energy spectral parameters. 
The low frequency QPOs are
related to the flux of the PL energy spectral component. The
variable frequency QPOs in XTE J1550-564 and GRO J1655-40 appear only
when the PL contributes more than $20\%$ of the $2-10\kev$ flux
(Sobczak et al. 2000a and references therein). GRS~1915+105 also shows
similar behavior (Muno et al. 1999).  The PL index is well
correlated with the centroid of the low frequency QPOs in GRS~1915+105
(Vignarca et al. 2003) and XTE~J1550+564 (Sobczak et al. 2000a,
2000b). The photon index increases with the QPO frequency until it
saturates. The frequency scaling factor between the low frequency QPOs
of two XRBs with similar PL photon indices can be used to
determine the ratio of their BH masses (TF04; FT04). Thus it is
crucial to determine the shape of X-ray spectrum and detect both the
QPO and the low frequency break in order to make sure that the
detected QPO lies at a  low frequency, but above the break frequency,
 and on top of a PDS power-law continuum.

The long \xmm{} observation of M82 X-1 has yielded all 3 pieces of
information required to determine its BH mass. The photon index of the
$3-12\kev$ PL is $2.0\pm0.1$ as determined from the high
signal-to-noise PN data. XRBs show a range of PL
photon indices from $\Gamma \sim 1.6$ in their low/hard state to
$\Gamma \sim 2.5$ in their high/soft state. Thus M82 X-1 is likely in
a state intermediate between the high and low states.  Sobczak et al.
(2000a) studied the spectral behavior of XTE~J1550-564 during its
1998--1999 outburst using 209 pointed observations with \rxte{}. They
used a model consisting of a multicolor disk blackbody and a PL
to fit the spectra and found that the spectra gradually steepen from
$\Gamma \sim 1.5$ to $\sim 2.9$ for XTE~J1550-564.  The low frequency
QPO is also found to vary from $0.08 - 13\hz$. At a PL photon
index of $1.91$ ($2.12$), XTE~J1550-564 shows a low frequency QPO at
$0.81\hz$ ($1.6\hz$) (Sobczak et al.  2000a; 2000b). Noting that the
BH mass scales inversely in proportion to the QPO frequency and the BH
mass of XTE~J1550-564 is about $10M_\odot$ (Titarchuk \& Shrader 2002,
Shrader \& Titarchuk 2003), we can determine the BH mass of the M82 X-1 
as
$M_{BH}$ (ULX) $\sim \nu_{QPO}$(XTE~J1550-564)/$\nu_{QPO}$(ULX)
$\times M_{BH}$ (XTE~J1550-564) $\sim 70 - 150 M\odot$. Another
XRB GRS~1915+105 shows a low frequency QPO with its centroid
at $0.642\pm0.004\hz$ ($1.609\pm0.007\hz$) when its PL photon
index is $1.88\pm0.04$ ($2.16\pm0.04$) (Vignarca et al. 2003).  Noting
that that the BH mass of GRS~1915+105 is $14\pm4 M\odot$ estimated
dynamically from the IR spectroscopic observation of its companion
star (Greiner et al. 2001), the BH mass for M82 X-1 is $M_{BH}$ (ULX)
$\sim 50 - 260M\odot$. Thus, scaling the mass of M~82 X-1 from the
 $\Gamma - \nu_{QPO}$ relation calibrated by the dynamical masses of
 several galactic BHs, we find $M_{BH}$ (ULX) $\sim 50 - 260M\odot$.
 
The QPO frequencies and photon indices for M82
X-1, XTE~J1550-564 and GRS~1915+105 correspond to a region of steep
dependence in the $\Gamma - \nu_{QPO}$ relation (see FT04).  It is
therefore likely that the uncertainty in the calibration of the $\Gamma -
\nu_{QPO}$ relation may introduce an additional factor of two
uncertainty in the determination of the BH mass of M~82~X-1.  Thus the
true BH mass of M~82 X-1 is most likely in the range of
$25-520M\odot$. This is at least a factor of $2$ lower than the
previous estimate by FT04, which can be explained by
the lower values of the QPO frequencies for M82 X-1 in our paper compared with
those in FT04.  The above estimate of the BH mass is based on the
assumption that M~82 X-1 also follows the $\Gamma - \nu_{QPO}$
relation  found for XRBs. \xmm{} and \rxte{}
observations of M~82 X-1 have revealed that the QPO frequency is
variable (SM03). The photon index is also variable (FT04) but appears
to be inconsistent with the $\Gamma - \nu_{QPO}$ relation for XRBs.
However, there may be systematic errors in the photon indices measured
with \xmm{} and \rxte{} due to contamination from nearby sources,
as indicated by large apparent changes in the effective absorption
column. \chandra{} ACIS observations of M~82 X-1 were used to suggest
 an effective
column in the range  $(5-9)\times10^{20}{\rm~cm^{-2}}$ (SM03), while
\xmm{} and \rxte{} observations have been used to suggest large columns
 in the range
$2\times10^{21} - 8\times10^{22}{\rm~cm^{-2}}$ (see Section 3.2,
FT04). Both \xmm{} and \rxte{} observations suffer from contamination
by nearby sources. A small contribution from nearby sources may
significantly alter the spectral shape of the ULX and a varying
contribution may lead to the observed changes in the absorption
column. Thus, based on the currently available X-ray data on M~82 X-1,
it is neither possible to prove nor rule out the $\Gamma - \nu_{QPO}$
relation similar to that found for XRBs. 
  
M82 X-1 is one of the brightest ULXs. During the second \xmm{}
observation, its $2-10\kev$  luminosity was $1.7\times
10^{40}{\rm~ergs~s^{-1}}$ (SM03), while it was a factor of $\sim 2$
brighter in the first \xmm{} observation.  Based on the first \xmm{}
observation, SM03 quoted a bolometric luminosity of $(4-5)\times
10^{40}{\rm~ergs~s^{-1}}$, equivalent to the Eddington luminosity for
a BH mass in the range of $300-400M\odot$. Given the uncertainty in
the contamination by nearby X-ray sources,  our estimated BH mass of M82~X-1
 is marginally consistent with the observed luminosity and the Eddington ratio.

\acknowledgements We thank an anonymous referee for  useful comments, Aspen center for Physics
for organizing the workshop
`Revealing black holes' and A. R. Rao for his comments.
GCD acknowledges the support of NASA grants through the
awards NNG04GN69G and NNG05GN35G.
This work is based on
observations obtained with \xmm{}, an ESA science mission with
instruments and contributions directly funded by ESA Member States and
the USA (NASA). 

\end{document}